\documentclass[runningheads, 11pt]{llncs}

\usepackage[letterpaper,top=2cm,bottom=2cm,left=3cm,right=3cm,marginparwidth=1.75cm]{geometry}

\usepackage{amsmath}
\usepackage{amssymb}
\usepackage{bm} % Bold maths
\usepackage{braket} % Dirac notation
\usepackage[super]{nth}
\usepackage[subdued,defaultmathsizes]{mathastext} % Maths in \texttt
\usepackage{import} % For easier pgf files
\usepackage{pgf}
\usepackage[font=small]{caption}
\usepackage{subcaption}
\usepackage[LGR, T1]{fontenc}
\usepackage{graphicx}
\usepackage[colorlinks=true, allcolors=blue]{hyperref}
\usepackage{titling} % Less space above title
\usepackage{authblk} % Affiliation
\usepackage[usestackEOL]{stackengine}
\usepackage{mathtools}
\usepackage{esvect} % Nicer vectors
\usepackage[section]{placeins} % Stop figures wandering into other sections
\usepackage{url}
\usepackage{float}

\usepackage[isbn=false]{biblatex}
\DeclareMathAlphabet{\mathgtt}{LGR}{cmtt}{m}{n}
\DeclareMathOperator*{\E}{\mathbb{E}}

\newenvironment{acknowledgements}
{\renewcommand\abstractname{Acknowledgements}\begin{abstract}} {\end{abstract}}

\title{Gradient Estimation with Constant Scaling for Hybrid Quantum Machine Learning}
\author{Thomas Hoffmann and Douglas Brown}
\affil{
Quantinuum, 17 Beaumont Street, Oxford, OX1 2NA, UK \\
\email{\{thomas.hoffmann, douglas.brown\}@quantinuum.com}
}

\begin{document}

\maketitle    

\begin{abstract}
We present a novel method for determining gradients of parameterised quantum circuits (PQCs) in hybrid quantum-classical machine learning models by applying the multivariate version of the  simultaneous perturbation stochastic approximation (SPSA) algorithm. The gradients of PQC layers can be calculated with an overhead of two evaluations per circuit per forward-pass independent of the number of circuit parameters, compared to the linear scaling of the parameter shift rule. These are then used in the backpropagation algorithm by applying the chain rule. We compare our method to the parameter shift rule for different circuit widths and batch sizes, and for a range of learning rates. We find that, as the number of qubits increases, our method converges significantly faster than the parameter shift rule and to a comparable accuracy, even when considering the optimal learning rate for each method.
\end{abstract}

\section{Introduction}
The field of quantum machine learning (QML) is a rapidly growing one. In the current NISQ (Noisy Intermediate Scale Quantum) era, the interplay of quantum and classical classical models appears to be an area of particular interest \cite{Schuld2019Quantum,FarhiClassification,McClean2016TheAlgorithms}. Indeed, there are many popular \texttt{python} packages available dedicated to facilitate the development of such hybrid models \cite{BroughtonTFQ,paszke2019pytorch,Wang2022GradientPruning,KartsaklisLambeq}. 

Optimisation is at the heart of all machine learning problems, quantum or otherwise. In the classical world, we use gradient-based optimisation methods that leverage a technique called \textit{reverse-mode Automatic Differentiation} (AD). AD allows the  gradients of computationally very heavy machine learning architectures to be calculated and enables the use of optimisation algorithms such as Stochastic Gradient Descent or \texttt{adam} \cite{Kiefer1952SGD,KingmaAdam}. 

Current quantum machine learning is typically performed using parameterised quantum circuits (PQCs), where trainable classical parameters are used to optimise the distribution of the measurement outcomes of a number of qubits. However, these quantum models come with the difficulty that the gradients of the output distributions with respect to the trainable input parameters to the PQC are not readily calculable due to the probabilistic nature of quantum mechanics. 

Two popular gradient-based methods of quantum circuits are the Simultaneous Perturbation Stochastic Approximation (SPSA) algorithm \cite{spall2000} and the Parameter Shift rule \cite{Mitarai2018QuantumLearning,Schuld2019Evaluating}. 
The former is a widely-used efficient optimisation method that estimates the gradient of a loss function by perturbing all model parameters simultaneously in random directions. This has the benefit that only two evaluations of the loss function are required per optimisation step, regardless of the dimension of the parameter space. 
A multivariate form has also been proposed to calculate Jacobian or Hessian matrices, but this has not yet been applied to quantum models \cite{Spall1992MultivariateApproximation}.

The latter is a more recent development specifically for PQCs which calculates analytical gradients using the same circuit with parameters that are shifted by multiples of $\pi/2$. However, the number of circuit evaluations required for the parameter shift rule scales linearly with the number of classical parameters, which can be undesirable for circuits of large width and depth.

In this work we apply the multivariate form of the the SPSA algorithm to the reverse-mode Automatic Differentiation algorithm to optimise quantum and hybrid models with constant scaling with respect to number of circuit parameters. We refer to our method as Simultaneous Perturbation for Stochastic Backpropagation (SPSB). In Section~\ref{sec:prev_work} we detail the background on Automatic Differentiation, the SPSA algorithm and quantum gradients. In Section~\ref{sec:spsb} we describe the method behind SPSB. Section~\ref{sec:experiments} details classification experiments carried out, the results of which are discussed in Section~\ref{sec:results}.

\section{Previous work} \label{sec:prev_work}
\subsection{Automatic Differentiation /  Backpropagation}\label{sec:ad}

Part of the success of Deep Learning architectures across many computer science domains is because of the use of gradient-based optimisation to find the optimal parameter setting within a space of possibly billions of dimensions. Prominent examples include algorithms like Stochastic Gradient Descent (SGD) or \texttt{adam} \cite{Kiefer1952SGD,KingmaAdam}. In order to calculate the required gradients, most Deep Learning frameworks like PyTorch or TensorFlow \cite{tf,paszke2019pytorch} leverage a technique called \textit{Reverse-mode Automatic Differentiation} (AD), nowadays mostly referred to as \textit{backpropagation} \cite{Griewank1989OnDifferentiation,Paszke2017AutomaticPytorch}.

Using AD, we record the computational graph of a model's forward pass and calculate the gradients of a loss function with respect to its parameters efficiently. This is done by subdividing the graph into the smallest possible entities that are easily algebraically differentiable and concatenating the partial derivatives using the chain rule. 

Consider a simple chained function $y=f(u(x))$, where $x\in \mathbb{R}^p$, $u: \mathbb{R}^p\rightarrow\mathbb{R}^m$ and $f: \mathbb{R}^m\rightarrow\mathbb{R}^n$. We start the backwards pass by computing $\partial y / \partial y$ which is simply a vector of 1s of size $\mathbb{R}^n$ which acts as a seed for the gradient \textit{upstream}. We now continue by calculating $\partial y / \partial u$, which is given by:
\begin{equation}
\label{eq:ad_1}
    \frac{\partial y}{\partial u} = \frac{\partial y}{\partial y}\frac{\partial y}{\partial u} = \frac{\partial y}{\partial y}\frac{\partial f}{\partial u}= \mathbf{\text{upstream}}\cdot\frac{\partial f}{\partial u},
\end{equation}
where $\partial f / \partial u$ is the Jacobian $\mathbf{J}_f$ with respect to $u$ (i.e. a matrix of size $\mathbb{R}^{n\times m}$). Treating the outcome of Eq.~\ref{eq:ad_1} as our new upstream vector ($\mathbf{\text{upstream}}\leftarrow \partial y / \partial u$), we calculate $\partial y / \partial x$ by:
\begin{equation}
    \frac{\partial y}{\partial x} = \frac{\partial y}{\partial u}\frac{\partial u}{\partial x} = \mathbf{\text{upstream}}\cdot\frac{\partial u}{\partial x},
\end{equation}
where $\partial u / \partial x$ is the Jacobian $\mathbf{J}_u$ with respect to $x$ (i.e. a matrix of size $\mathbb{R}^{m\times p}$).

We can repeat this procedure for an arbitrary number of inner functions and, most importantly, decompose any differentiable function $y=f(x)$ into a chain of elementary operations whose derivatives we know exactly. The parameters of the model can then be updated using an optimiser, which generally takes a form:
\begin{equation}
    x_{n+1} = x_n - \alpha \left.\frac{\partial f}{\partial x}\right\vert_{x=x_n},
\end{equation}
where $\alpha$ is a (possibly variable) learning rate determined by the optimiser used.

While reverse-mode AD is suitable for classical computations, it is not possible to backpropagate through quantum circuits in the same way as we don't have access to the upstream variable due to the unobservable nature of the intermediate quantum states.

\subsection{Quantum gradients}
The parameter shift rule \cite{Mitarai2018QuantumLearning,Schuld2019Evaluating} is a widely used method of optimising the parameters of variational quantum circuits \cite{Benedetti2019ParameterizedModels,Cerezo2021VariationalAlgorithms}. It was developed with a particular focus on evaluating gradients on real quantum hardware, where it is desirable to calculate gradients using the same circuit structure. 

Consider a unitary parameterised by a single angle, $U(\theta)$, acting on a quantum circuit, and a function of the circuit defined as:
\begin{equation}
    f(\theta) := \braket{\psi | U^\dagger (\theta) \, O \, U(\theta) | \psi} = \braket{\psi | \mathcal{M}(\theta) | \psi},
\end{equation}
for some state $\psi$ and observable $O$. Using the parameter shift rule, the gradient of this function with respect to the angle $\theta$, $\nabla_\theta$, can be expressed as a linear combination of $\mathcal{M}(\theta \pm \Delta \theta)$, for some finite (i.e. non infinitesimal) $\Delta \theta$:
\begin{equation} \label{eq:param_shift}
    \nabla_\theta \mathcal{M}(\theta) = c \left [ \mathcal{M}(\theta + \Delta \theta) - \mathcal{M}(\theta - \Delta \theta) \right].
\end{equation} 
Parameters $c$ and $\Delta \theta$ are determined by the unitary and measurements in question.

For larger circuits with more than one parameter, the number of circuit executions required for the parameter shift rule scales as $\mathcal{O}(n)$ where $n$ is the number of parameters to be optimised, as the gradient must be taken with respect to each parameter individually.

Various works have attempted to optimise the effiency with which the parameter shift rule converges. These include probabilistic gradient pruning to remove gradients which are likely to have higher errors \cite{Wang2022GradientPruning}, and the quantum natural gradient \cite{Stokes2020NaturalGradient}, which uses the Fisher information matrix to help finding the path of steepest descent in the quantum landscape. The SPSA algorithm has also been used to calculate this Fisher information matrix, but has not been used to calculate the gradients directly \cite{Gacon2021SimultaneousInformation}.

Other tools have been developed for optimisation of quantum circuits, such as the gradient-free rotosolve algorithm, and the application of the adjoint state method for gradients of quantum simulations \cite{Luo2020Yao.jl:Design,Ostaszewski2019StructureCircuits,Plessix2006AApplications}. However, none of these methods are capable of backpropagation through hybrid models when using real quantum hardware.

\subsection{The SPSA Algorithm}
The Simultaneous Perturbation Stochastic Approximation algorithm is a powerful optimisation method used to find solutions to optimisation problems of the form \cite{Spall1992MultivariateApproximation}:
\begin{equation}
    \min_\mathbf{x} f(\mathbf{x}) = \min_\mathbf{x} 
    \E_{\xi}\left[F(\mathbf{x}, \xi)\right].
\end{equation}
Here, $F$ is a function $\mathbb{R}^n \rightarrow \mathbb{R}$ which takes a set of parameters $\mathbf{x} \in \mathbb{R}^n$ and is subject to a noise term $\xi$. The aim is to find $\mathbf{x}$ such that the expectation value of $F(\mathbf{x}, \xi)$ is minimal.

To use a gradient-based optimisation scheme, we estimate $\nabla f(\mathbf{x})$ by perturbing all parameters using a randomly initialised vector $\mathbf{\Delta}$ with elements $\Delta_i$, defined as:
\begin{equation}
    \Delta_i = 
    \begin{cases}
    -1 & \textrm{with probability 0.5}\\
    1 & \textrm{otherwise}
    \end{cases}
\end{equation}
where $0 \leq i < n$. Fixing a small grid size $\epsilon$, we now calculate:
\begin{align}
    &&&&f_+ = f(\mathbf{x} + \epsilon \cdot \mathbf{\Delta}) && \text{and} && f_- = f(\mathbf{x} - \epsilon \cdot \mathbf{\Delta}).&&&&
\end{align}
This allows us to approximate the gradient $\nabla f|_{\mathbf{x}}$ as follows:
\begin{equation} \label{eq:grad_approx}
    \nabla f|_{\mathbf{x}} \approx \frac{f_+ - f_-}{2 \epsilon \mathbf{\Delta}}.
\end{equation}
We then update the model parameters via $\mathbf{x}_{n+1} = \mathbf{x}_n - \alpha \nabla \left.f\right\vert_{\mathbf{x}_n}$,
where $\alpha$ is a small learning rate. Using a sufficient number of iterations, the algorithm will approach the desired minimum efficiently. Vanilla SPSA implements grid size ($\epsilon$) decay and learning rate ($\alpha$) decay which we omit here for simplicity. Furthermore, we are only interested in finding the gradient, rather than using the full SPSA optimisation algorithm.

\section{SPSA for Backpropagation} \label{sec:spsb}
\subsection{Method}
 In order to perform backpropagation through our quantum (or hybrid) model, we need to approximate the Jacobian $J$ of a (potentially noisy) multi-variate function $\bm{f}: \mathbb{R}^n \rightarrow \mathbb{R}^m$:
\begin{equation}
    J =
    \begin{bmatrix}
    \frac{\partial \bm{f}}{\partial x_1} & \cdots & \frac{\partial \bm{f}}{\partial x_n}
    \end{bmatrix}
    = 
    \begin{bmatrix}
    \nabla^T f_1 \\
    \vdots\\
    \nabla^T f_m
    \end{bmatrix}
    =
    \begin{bmatrix}
    \frac{\partial f_1}{\partial x_1} & \cdots & \frac{\partial f_1}{\partial x_n} \\
    \vdots & \ddots & \vdots \\
    \frac{\partial f_m}{\partial x_1} & \cdots & \frac{\partial f_m}{\partial x_n}
    \end{bmatrix}.
\end{equation}

The $i$th row of $J$ corresponds to the vector $\nabla^T f_i$ which we can approximate with the SPSA rule above (Eq. \ref{eq:grad_approx}):
\begin{equation}
    \nabla f_i|_{\mathbf{x}} \approx \frac{f_{i, +} - f_{i, -}}{2 \epsilon \mathbf{\Delta}}.
\end{equation}
It is now easy to verify that, given a random vector $\mathbf{\Delta}$ and a grid parameter $\epsilon$, we can estimate the Jacobian using:
\begin{equation}
    J \approx \frac{1}{2\epsilon} \cdot (\bm{f}_{+} - \bm{f}_{-}) \otimes \mathbf{\Delta}^{\odot -1} \kern-16pt,
\end{equation}
where we denote by $\otimes$ the outer product, and by $\mathbf{\Delta}^{\odot -1}$ the element-wise inverse of $\mathbf{\Delta}$. We remark that in Ref. \cite{spall2000}, this method is used to estimate the Hessian of a loss function (i.e. the Jacobian of the gradient) for a second-order Newton-Raphson optimisation method.

Now that we have an estimation of the Jacobian matrix, we can use SPSA-like gradients for Automatic Differentiation as described in Section~\ref{sec:ad}. This method, which we refer to as SPSB, was implemented in PennyLane, a toolbox to develop and test classical-quantum hybrid models \cite{PennyLane}. Furthermore, we provide SPSB implementations for TensorFlow-Quantum \cite{BroughtonTFQ} and qujax \cite{qujax2022}.

\subsection{Convergence analysis}

Given a data set $\mathcal{D}=\{x_i\}^n$ and a loss function of a parameterised predictive model, $l(x, \boldsymbol\theta)$, we define 
$$\bar{l}(\boldsymbol\theta)=\frac{1}{|\mathcal{D}|}\sum^{|\mathcal{D}|}_{i=1} l(x_i,\boldsymbol\theta)$$
to be the global average of $l$, given a data set $\mathcal{D}$.

The general optimisation objective is to find $\boldsymbol\theta^*=\min_\theta \bar{l}(\boldsymbol\theta)$. In most machine learning use cases, evaluating $\bar{l}$ is not feasible as $|\mathcal{D}|$ is too large. Therefore, we estimate the global gradient $\partial\bar{l} /\partial\boldsymbol\theta$ using a batch $b$ of data points $\{x_i\}^b \in \mathcal{D}$. We easily see that $\hat{l}_i(\boldsymbol\theta)=l(x_i,\boldsymbol\theta)$ is an unbiased estimator of $\bar{l}$. Further, the gradient,
\begin{equation}
    \hat{g}_i = \frac{\partial \hat{l}_i(\boldsymbol\theta)}{\partial \boldsymbol\theta},
\end{equation}
is an unbiased estimator of the global gradient \mbox{$g=\partial\bar{l}(\boldsymbol\theta)/\partial\theta$}. Averaging the estimated gradients using $b$ samples yields \mbox{$\hat{g}^{(b)} = \frac{1}{b} \sum^b_{i=1} \hat{g}_i$}, which is used for the parameter update \mbox{$\boldsymbol\theta_{n+1}=\boldsymbol\theta_n-\alpha\hat{g}^{(b)}$}. Intuitively, we note that increasing the batch size lowers the variance of $\hat{g}^{(b)}$, i.e. stabilising the parameter update. However, a certain level of noise is desirable to escape local minima in the parameter landscape \cite{Duffield2022Bayesian}.

In the use case of SPSB, for each element in $\{x_n\}^b$, we record the forward pass $\hat{l}(x_n, \boldsymbol\theta)$ and the perturbed $\hat{l}^+_n=l(x_n, \boldsymbol\theta+\epsilon\boldsymbol\Delta)$ and $\hat{l}^-_n=l(x_n, \boldsymbol\theta-\epsilon\boldsymbol\Delta)$. As previously seen, we can estimate the gradient of $l$ with respect to $\boldsymbol\theta$ with

\begin{equation}
    \tilde{g}_n = \frac{\hat{l}^+_n-\hat{l}^-_n}{2\epsilon\boldsymbol\Delta}.
\end{equation}

According to \cite{spall2000}, we observe that $\mathbb{E}[\tilde{g}_n]=\hat{g}_n+\mathcal{O}(\epsilon^2)$, i.e. $\tilde{g}_n$ is an almost unbiased estimator of $\hat{g}$ if $\epsilon \rightarrow 0$ through the optimisation process. As we are considering only a single gradient evaluation, our gradient has a bias of $\mathcal{O}(\epsilon^2)$.
Given that the sampling variance of $\tilde{g}_n$ is high, we note that a larger batch size $b$ helps to reduce the stochastic noise in the optimisation step, yielding a better estimate of $\boldsymbol\theta^*$.

It is also worth considering that what we present is not an optimisation algorithm, and that the primary proposed usage of SPSB is for hybrid models where the gradient of the quantum module is only part of the global gradient. We therefore expect this small bias in the gradient not to substantially affect the convergence of a model. Nevertheless, we note that it desirable to choose a very small $\epsilon$, which depends on the resolution of the quantum computer or underlying (noisy) simulator. For the following experiments we choose a constant shift parameter of $\epsilon=0.01$.

\section{Experiments} \label{sec:experiments}

\subsection{Random datapoints benchmark}
\label{sec:rand_dps}

To test SPSB, we record the loss of QML models being trained on a classification task using randomly initialised datapoints. We use $N_q$ qubits and a batch size of $N_B$, and two different models, as illustrated below. One is a purely quantum model with a single qubit expectation used to classify datapoints, and the other feeds all the qubit expectations into a classical fully connected (FC) layer before calculating the loss.

\[\Lstackgap = 10pt\texttt{input} 
\rightarrow 
\texttt{PQC} \,  % j is \theta in \mathgtt{}
\Vectorstack{%
\nearrow \\%[-1.5ex]
 \searrow }
\:
\begin{matrix*}[l]
 \texttt{one-qubit expectation} 
 \rightarrow
 \texttt{loss}
 \hspace{10pt} (a)\\[3ex]
 \texttt{all-qubit expectations}
 \rightarrow
 \texttt{FC layer}
 \rightarrow
 \texttt{loss}
 \hspace{10pt} (b)
\end{matrix*} \]

We generate 100 datapoints for a binary classification task with $N_f$ features  equal to the number of qubits, $N_q$. We use angle encoding to input these into a layer of $R_X$ gates, followed by an Instantaneous Quantum Polynomial (IQP) ansatz with $N_l$ layers for classification \cite{Shepherd2009TemporallyComputation}. $N_l$ is 3 for all experiments here. The total number of trainable parameters in the circuit is equal to $N_q \times N_l$. All qubits are measured in the $Z$ basis. Binary cross entropy is used as the loss function and \texttt{adam} \cite{KingmaAdam} as the optimiser with a learning rate of 0.01. The batch size used is 25. Note that we don't split the data set into train and validation, as we are purely interested in the convergence behaviour of our algorithm, which can be ideally observed by intentionally overfitting the QML models to the data.

\subsection{MNIST Quanvolutional Network}

\begin{figure}[bth]
    \centering
    \includegraphics[width=0.7\textwidth]{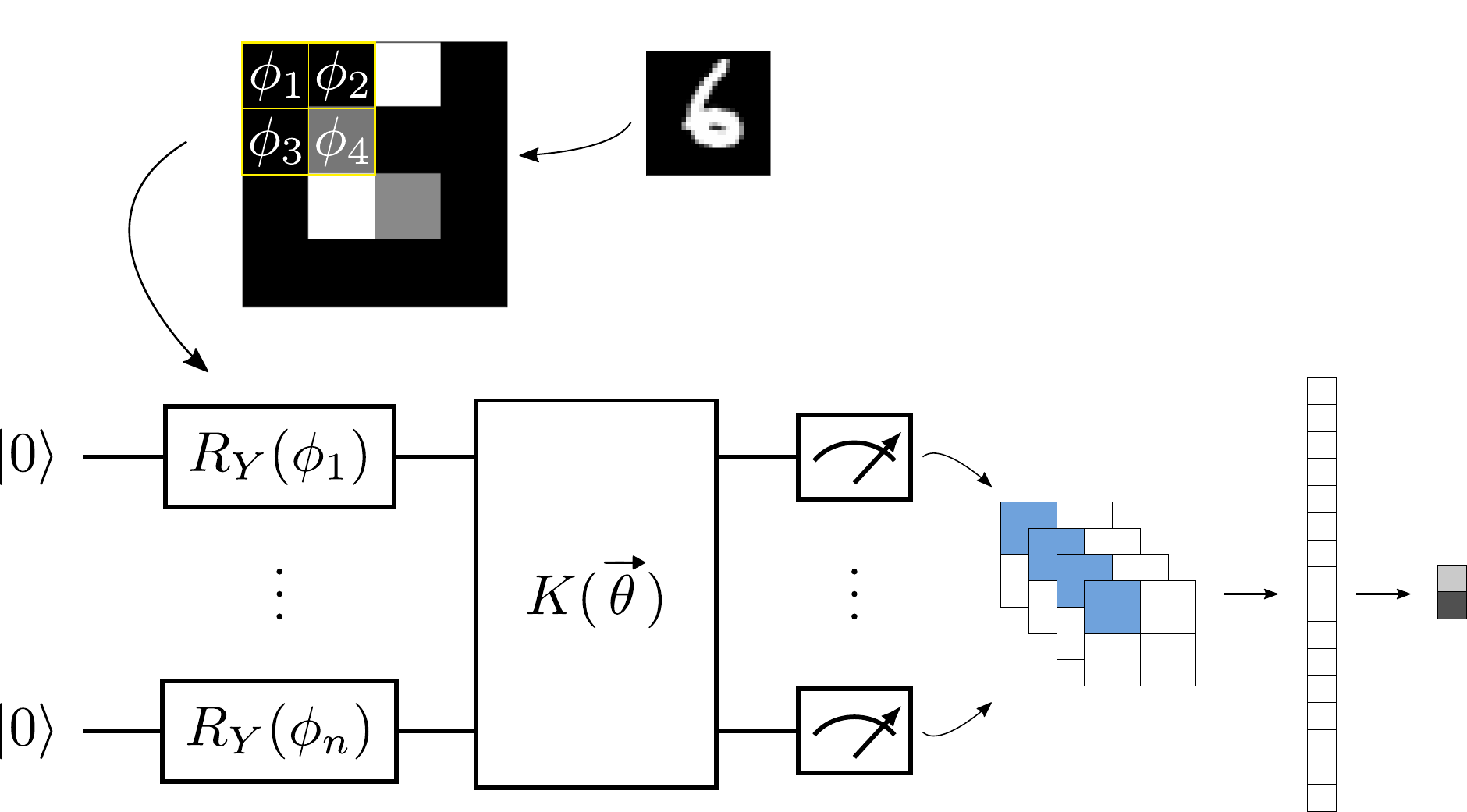}
    \caption{Diagram illustrating a so-called `quonvolutional' hybrid network, first introduced in Ref. \cite{Henderson2020QuanvolutionalCircuits}.}
    \label{fig:quonv}
\end{figure}

Here, we demonstrate a more practical application of the SPSB algorithm. We consider a so-called `quanvolutional neural network' \cite{Henderson2020QuanvolutionalCircuits}, inspired by the ubiquitous convolutional networks used in classical neural image processing.

The setup is illustrated in Fig. \ref{fig:quonv}. An input image is scanned over by a window of a certain size. For each window-sized section of image, the (normalised) pixel values are fed into a PQC using angle encoding. The number of qubits is equal to the area of the window. A variational `kernel' ansatz with trainable parameters $\mathbf{\theta}$ is then used to produce measurements on each of the qubits, and these are fed into a neural network to provide the classification. (In the original work a random circuit is used for the kernel ansatz -- here we use a trainable kernel to illustrate the use of SPSB).

To test the model we use the MNIST handwritten image dataset \cite{Lecun1998Gradient-basedRecognition}, resized from \mbox{28$\times$28} pixels to \mbox{4$\times$4}. The window size used is \mbox{2$\times$2}, meaning that 4 qubits are used. The ansatz used is consist of 3 layers of an IQP ansatz.
The output expectations are concatenated, flattened, and a linear neural network layer with softmax activation is used to predict the classification. The dataset is filtered to include only images of 6s and 3s, and a total of 1000 images are used. The batch size used is 50 and the model is trained with the \texttt{adam} optimiser.

\section{Results and Discussion} \label{sec:results}

\begin{figure}[H]
    \captionsetup[subfigure]{justification=justified,margin=1cm,width=4cm,singlelinecheck=false}
    \begin{subfigure}{0.43\textwidth}
        \includegraphics[]{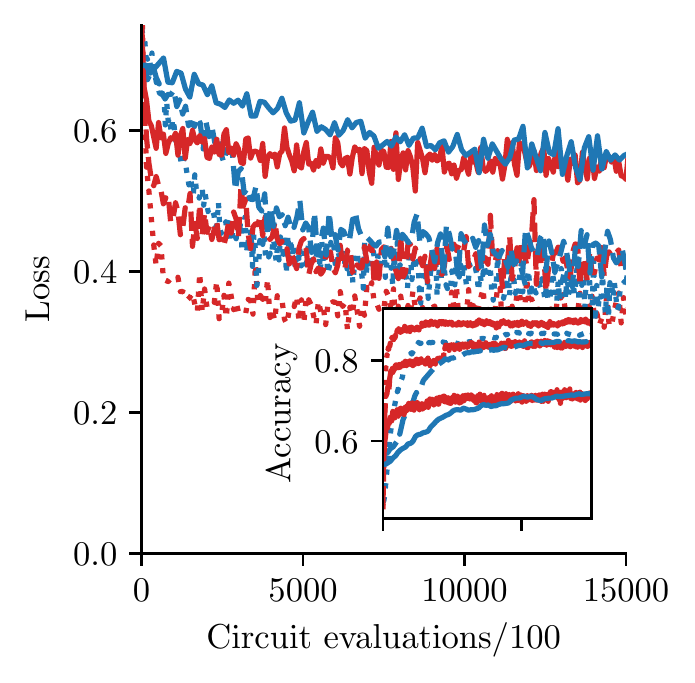}
        \caption{$\texttt{input}\rightarrow\texttt{PQC}\rightarrow\texttt{loss}$}
        \label{fig:spsa_v_paramshift:a}
    \end{subfigure}
    \captionsetup[subfigure]{justification=justified,margin=0.2cm,singlelinecheck=false}
    \begin{subfigure}{0.57\textwidth}
        \includegraphics[]{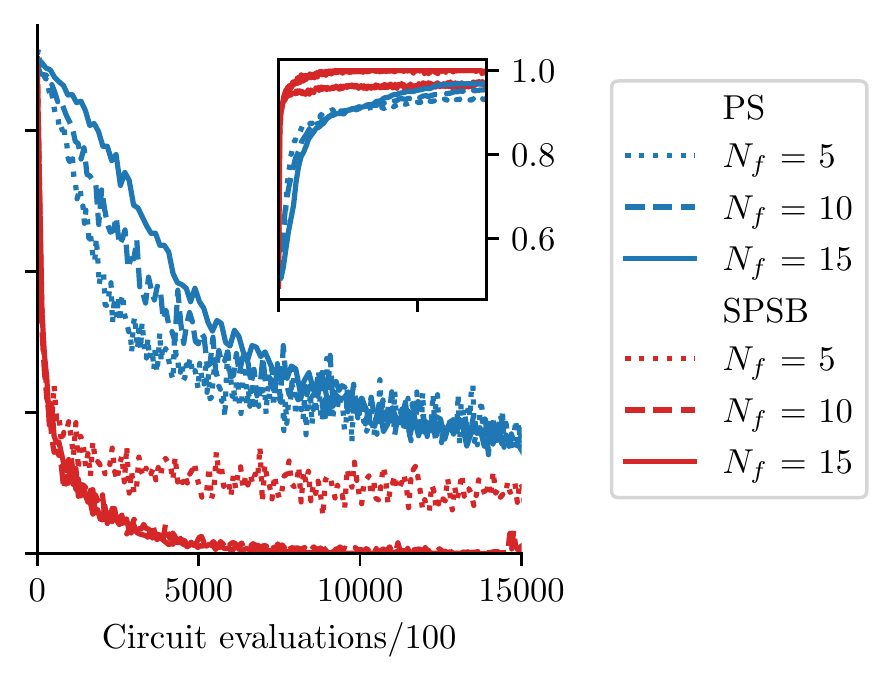}
        \caption{$\texttt{input}\rightarrow\texttt{PQC}\rightarrow\texttt{FC}\rightarrow\texttt{loss}$}
        \label{fig:spsa_v_paramshift:b}
    \end{subfigure}
    \caption{Figures comparing training loss and accuracy using gradients from the parameter shift rule (blue) and the SPSB algorithm (red). Figure (a) shows a quantum-only and (b) a quantum-classical hybrid model. Each line shows a different number of features $N_f$ of the data, which is set to be equal to the number of qubits $N_q$. We use a batch size of 25 for training. Every line is the median of 10 independent runs. For better readability, we show the loss after every 100th optimisation step for SPSB and every 3rd step for the parameter shift rule.}
    \label{fig:spsa_v_paramshift}
\end{figure}

\begin{figure}[H]
    \captionsetup[subfigure]{justification=justified,margin=0.5\columnwidth,width=5cm,singlelinecheck=false}
    \begin{subfigure}{0.43\textwidth}
    \includegraphics{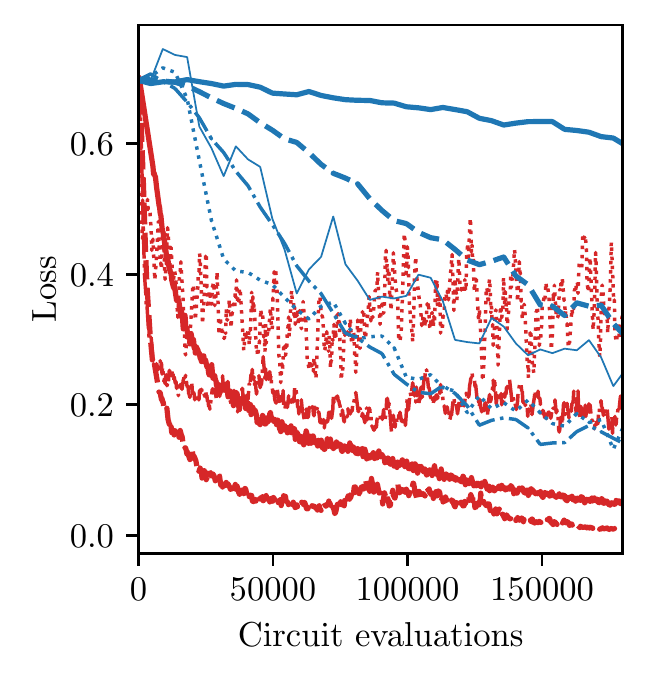}
        \caption{Random data task, $N_B=25$}
        \label{fig:lrs:a}
    \end{subfigure}
    \captionsetup[subfigure]{justification=justified,margin=0.8cm,singlelinecheck=false}
    \begin{subfigure}{0.57\textwidth}
        \includegraphics{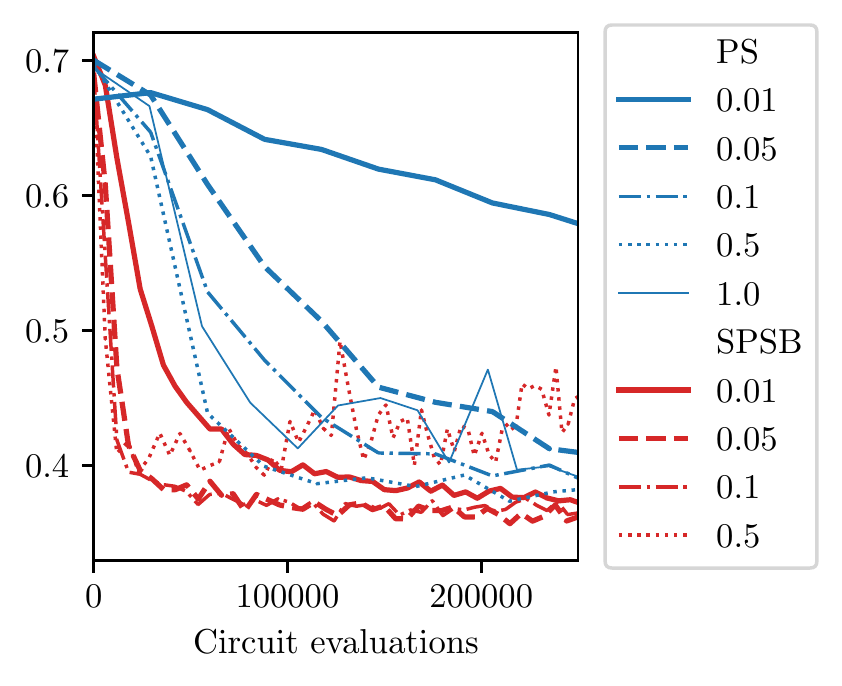}
        \caption{`Quanvolutional' MNIST task, $N_B=50$}
        \label{fig:lrs:b}
    \end{subfigure}
    \caption{Figures showing training loss for the parameter shift rule for two different hybrid tasks and a range of learning rates. To enhance readability of the diagram, we record the rolling averages of the loss curves, with a window size of 3 for the parameter shift and 10 for SPSB. Figure (a) shows the random datapoints task with 15 features and a batch size of 25. Each line is the median of 10 iterations. Figure (b) shows the quanvolutional network with 4 qubits, as described above, and a batch size of 50. Each line is the median of 5 iterations. For transparency, the raw data without taking the rolling averages is shown in the \hyperref[sec:appendix]{Appendix}.}
    \label{fig:learning_rates}
\end{figure}

Figure \ref{fig:spsa_v_paramshift} shows the loss and accuracy of the parameter shift rule and SPSB on the random dataset task described in Sec.~\ref{sec:rand_dps} for a range of feature/qubit numbers, $N_f$. The scaling benefit of the SPSB algorithm compared to the parameter shift rule is clear from the rapid decrease of the loss function for higher numbers of features in comparison to the parameter shift rule. For a higher learning rate of 0.1 (c.f. Fig.~\ref{fig:lrs:a}) we observed noise and less stable convergence in the SPSB data for higher $N_f$, whereas the parameter shift rule seemed to cope better with the higher learning rate.  A small but inconsequential noise artefact is present in Fig.~\ref{fig:spsa_v_paramshift:b}.

Despite the favourable performance of the SPSB algorithm in this example, we find that for very low batch sizes the global minimum is not always found, while the performance using parameter-shift gradients is unaffected by the batch size. We hypothesise that this is because the gradients are averaged over all samples in the batch, and so for very small batch sizes the random variation inherent in the algorithm prevents it from descending the landscape correctly.

We consider the effect of learning rate on the convergence behaviour of SPSB and the parameter shift rule. The results of this are shown in Fig.~\ref{fig:learning_rates}. We consider each of the tasks above, with batch sizes of 25 and 50 respectively, and the \texttt{adam} optimiser with a given learning rate. The figures show the median behaviour of 10 and 5 iterations respectively, and a rolling average is taken to aid in clarity, since the original plots are rather noisy. In part this may be due to the stochastic nature of SPSB, but also, because we compare the convergence over the number of circuit evaluations, we record many more parameter updates for SPSB than for the parameter shift rule. For transparency, the original plots are shown in the \hyperref[sec:appendix]{Appendix}.

In Fig.~\ref{fig:lrs:a} we observe that the rate of convergence is affected substantially by the learning rate. The rate of convergence of the parameter shift rule increases with increasing learning rate from 0.01 to 0.5. SPSB appears to prefer a lower learning rate, with $lr = 0.05$ being the best of those tested on the random data set task. A higher learning rate of 0.5 causes SPSB not to converge, presumably due to excess statistical noise. Comparing the optimal learning rate for the parameter shift rule ($lr=0.5$) with the optimal learning rate for SPSB ($lr=0.01$), we find that SPSB requires up to 90\% fewer circuit evaluations to reach the same loss.

For the `quanvolutional' model, shown in Fig.~\ref{fig:lrs:b}, SPSB converges significantly quicker than the parameter shift rule, even considering each method's optimal learning rate. The effect of learning rate on the convergence in this task is similar to Fig.~\ref{fig:lrs:a}. Learning rates of 0.05 and 0.1 perform very similarly with SPSB, while 0.5 is clearly too large. The parameter shift rule appears to have an optimal learning rate for this task between 0.5 and 1.0. The advantage presented here will become clearer still with a larger model with a deeper ansatz or a larger window size.

\section{Conclusion} \label{sec:conclusion}
The key contribution of this work is a simple, general-purpose method for gradient estimation which significantly improves the speed of training quantum hybrid models due to its favourable scaling over the parameter shift rule. This is particularly valuable for training on real NISQ hardware (and of course simulation thereof). We show the performance of the algorithm on a random datapoints classification task for a range of model sizes and hyperparameters, and on a `quanvolutional' hybrid model. We examine the effect of learning rate on the two gradient methods and demonstrate the advantage of SPSB over the parameter shift rule even for their respective optimal learning rates.

We show that this algorithm is well-adapted to cope with large batches of data and with circuits with large parameter numbers, which will become ever more important as quantum hardware develops over the coming years. The method presented may have difficulties with small batch sizes, or tasks where batching is not feasible such as combinatorial optimisation, but this can be mitigated by averaging the Jacobian over several evaluations, as suggested in Ref. \cite{spall2000}. For large batch sizes, we expect our method to help avoid local minima due to the introduction of additional statistical noise on the gradients which can help escape regions in the parameter landscape with zero-gradients, i.e. barren plateaus.

For the end user, taking advantage of SPSB is straightforward, especially for libraries which by design allow a choice of differentiator. Our git repository (linked below) contains implementations for some of the major quantum machine learning libraries. 

There are also a number of avenues for further study. The SPSA optimiser is typically used with stochastic gradient descent, but this work opens up the possibility of its use with other optimisers such as \texttt{Adagrad} and \texttt{adam}, even when considering purely quantum models such as VQE. The quantum natural gradient method \cite{Stokes2020NaturalGradient} could also be applied while using SPSB to better track the gradient down the loss landscape.

The code used for this work is available \href{https://github.com/CQCL/Constant-Scaling-Quantum-Gradient-Estimation}{here}.

\begin{acknowledgements}
The authors are grateful to Konstantinos Meichanetzidis, Sam Duffield, and Marcello Benedetti for helpful discussion and feedback during the course of this work.
\end{acknowledgements}

\printbibliography

@software{qujax2022,
  author = {Samuel Duffield and Kirill Plekhanov and Gabriel Matos and Melf Johannsen},
  title = {qujax: Simulating quantum circuits with JAX},
  url = {https://github.com/CQCL/qujax},
  year = {2022},
}

@misc{Duffield2022Bayesian,
  doi = {10.48550/ARXIV.2206.07559},
  url = {https://arxiv.org/abs/2206.07559},
  author = {Duffield, Samuel and Benedetti, Marcello and Rosenkranz, Matthias},
  title = {Bayesian Learning of Parameterised Quantum Circuits},
  publisher = {arXiv},
  year = {2022},
  copyright = {arXiv.org perpetual, non-exclusive license}
}

@article{paszke2019pytorch,
  title={Pytorch: An imperative style, high-performance deep learning library},
  author={Paszke, Adam and Gross, Sam and Massa, Francisco and Lerer, Adam and Bradbury, James and Chanan, Gregory and Killeen, Trevor and Lin, Zeming and Gimelshein, Natalia and Antiga, Luca and others},
  journal={Advances in neural information processing systems},
  volume={32},
  year={2019}
}

@misc{KingmaAdam,
    title = {{Adam: A Method for Stochastic Optimization}},
    year = {2014},
    author = {Kingma, Diederik P and Ba, Jimmy},
    publisher = {arXiv},
    url = {https://arxiv.org/abs/1412.6980},
    doi = {10.48550/ARXIV.1412.6980}
}

@inproceedings{spall2000,
    title = {{Adaptive stochastic approximation by the simultaneous perturbation method}},
    booktitle = {Proceedings of the 37th IEEE Conference on Decision and Control (Cat. No.98CH36171)},
    author = {Spall, J.C.},
    pages = {3872--3879},
    publisher = {IEEE},
    isbn = {0-7803-4394-8},
    doi = {10.1109/CDC.1998.761833}
}

@misc{FarhiClassification,
    title = {{Classification with Quantum Neural Networks on Near Term Processors}},
    year = {2018},
    author = {Farhi, Edward and Neven, Hartmut},
    publisher = {arXiv},
    url = {https://arxiv.org/abs/1802.06002},
    doi = {10.48550/ARXIV.1802.06002}
}

@article{Schuld2019Evaluating,
    title = {{Evaluating analytic gradients on quantum hardware}},
    year = {2019},
    journal = {Physical Review A},
    author = {Schuld, Maria and Bergholm, Ville and Gogolin, Christian and Izaac, Josh and Killoran, Nathan},
    number = {3},
    month = {3},
    pages = {032331},
    volume = {99},
    url = {https://link.aps.org/doi/10.1103/PhysRevA.99.032331},
    doi = {10.1103/PhysRevA.99.032331},
    issn = {2469-9926}
}

@misc{KartsaklisLambeq,
    title = {{lambeq: An Efficient High-Level Python Library for Quantum NLP}},
    year = {2021},
    author = {Kartsaklis, Dimitri and Fan, Ian and Yeung, Richie and Pearson, Anna and Lorenz, Robin and Toumi, Alexis and de Felice, Giovanni and Meichanetzidis, Konstantinos and Clark, Stephen and Coecke, Bob},
    publisher = {arXiv},
    url = {https://arxiv.org/abs/2110.04236},
    doi = {10.48550/ARXIV.2110.04236}
}

@misc{PennyLane,
    title = {{PennyLane: Automatic differentiation of hybrid quantum-classical computations}},
    year = {2018},
    author = {Bergholm, Ville and Izaac, Josh and Schuld, Maria and Gogolin, Christian and Ahmed, Shahnawaz and Ajith, Vishnu and Alam, M Sohaib and Alonso-Linaje, Guillermo and AkashNarayanan, B and Asadi, Ali and Arrazola, Juan Miguel and Azad, Utkarsh and Banning, Sam and Blank, Carsten and Bromley, Thomas R and Cordier, Benjamin A and Ceroni, Jack and Delgado, Alain and Di Matteo, Olivia and Dusko, Amintor and Garg, Tanya and Guala, Diego and Hayes, Anthony and Hill, Ryan and Ijaz, Aroosa and Isacsson, Theodor and Ittah, David and Jahangiri, Soran and Jain, Prateek and Jiang, Edward and Khandelwal, Ankit and Kottmann, Korbinian and Lang, Robert A and Lee, Christina and Loke, Thomas and Lowe, Angus and McKiernan, Keri and Meyer, Johannes Jakob and Monta{\~{n}}ez-Barrera, J A and Moyard, Romain and Niu, Zeyue and O'Riordan, Lee James and Oud, Steven and Panigrahi, Ashish and Park, Chae-Yeun and Polatajko, Daniel and Quesada, Nicolás and Roberts, Chase and S{\'{a}}, Nahum and Schoch, Isidor and Shi, Borun and Shu, Shuli and Sim, Sukin and Singh, Arshpreet and Strandberg, Ingrid and Soni, Jay and Sz{\'{a}}va, Antal and Thabet, Slimane and Vargas-Hern{\'{a}}ndez, Rodrigo A and Vincent, Trevor and Vitucci, Nicola and Weber, Maurice and Wierichs, David and Wiersema, Roeland and Willmann, Moritz and Wong, Vincent and Zhang, Shaoming and Killoran, Nathan},
    publisher = {arXiv},
    url = {https://arxiv.org/abs/1811.04968},
    doi = {10.48550/ARXIV.1811.04968}
}

@inproceedings{Wang2022GradientPruning,
    title = {{QOC}},
    year = {2022},
    booktitle = {Proceedings of the 59th ACM/IEEE Design Automation Conference},
    author = {Wang, Hanrui and Li, Zirui and Gu, Jiaqi and Ding, Yongshan and Pan, David Z. and Han, Song},
    month = {7},
    pages = {655--660},
    publisher = {ACM},
    address = {New York, NY, USA},
    isbn = {9781450391429},
    doi = {10.1145/3489517.3530495}
}

@article{Schuld2019Quantum,
    title = {{Quantum Machine Learning in Feature Hilbert Spaces}},
    year = {2019},
    journal = {Physical Review Letters},
    author = {Schuld, Maria and Killoran, Nathan},
    number = {4},
    month = {2},
    pages = {040504},
    volume = {122},
    doi = {10.1103/PhysRevLett.122.040504},
    issn = {0031-9007}
}

@article{Stokes2020NaturalGradient,
    title = {{Quantum Natural Gradient}},
    year = {2020},
    journal = {Quantum},
    author = {Stokes, James and Izaac, Josh and Killoran, Nathan and Carleo, Giuseppe},
    month = {5},
    pages = {269},
    volume = {4},
    doi = {10.22331/q-2020-05-25-269},
    issn = {2521-327X}
}

@article{Kiefer1952SGD,
    title = {{Stochastic Estimation of the Maximum of a Regression Function}},
    year = {1952},
    journal = {The Annals of Mathematical Statistics},
    author = {Kiefer, J and Wolfowitz, J},
    number = {3},
    pages = {462 – 466},
    volume = {23},
    publisher = {Institute of Mathematical Statistics},
    url = {https://doi.org/10.1214/aoms/1177729392},
    doi = {10.1214/aoms/1177729392}
}

@misc{BroughtonTFQ,
    title = {{TensorFlow Quantum: A Software Framework for Quantum Machine Learning}},
    year = {2020},
    author = {Broughton, Michael and Verdon, Guillaume and McCourt, Trevor and Martinez, Antonio J and Yoo, Jae Hyeon and Isakov, Sergei V and Massey, Philip and Halavati, Ramin and Niu, Murphy Yuezhen and Zlokapa, Alexander and Peters, Evan and Lockwood, Owen and Skolik, Andrea and Jerbi, Sofiene and Dunjko, Vedran and Leib, Martin and Streif, Michael and Von Dollen, David and Chen, Hongxiang and Cao, Shuxiang and Wiersema, Roeland and Huang, Hsin-Yuan and McClean, Jarrod R and Babbush, Ryan and Boixo, Sergio and Bacon, Dave and Ho, Alan K and Neven, Hartmut and Mohseni, Masoud},
    publisher = {arXiv},
    url = {https://arxiv.org/abs/2003.02989},
    doi = {10.48550/ARXIV.2003.02989}
}

@inproceedings{tf,
    title = {{TensorFlow: A System for Large-Scale Machine Learning}},
    year = {2016},
    booktitle = {Proceedings of the 12th USENIX Conference on Operating Systems Design and Implementation},
    author = {Abadi, Martín and Barham, Paul and Chen, Jianmin and Chen, Zhifeng and Davis, Andy and Dean, Jeffrey and Devin, Matthieu and Ghemawat, Sanjay and Irving, Geoffrey and Isard, Michael and Kudlur, Manjunath and Levenberg, Josh and Monga, Rajat and Moore, Sherry and Murray, Derek G and Steiner, Benoit and Tucker, Paul and Vasudevan, Vijay and Warden, Pete and Wicke, Martin and Yu, Yuan and Zheng, Xiaoqiang},
    pages = {265--283},
    series = {OSDI'16},
    publisher = {USENIX Association},
    address = {USA},
    isbn = {9781931971331}
}

@article{Plessix2006AApplications,
    title = {{A review of the adjoint-state method for computing the gradient of a functional with geophysical applications}},
    year = {2006},
    journal = {Geophysical Journal International},
    author = {Plessix, R.-E.},
    number = {2},
    month = {11},
    pages = {495--503},
    volume = {167},
    doi = {10.1111/j.1365-246X.2006.02978.x},
    issn = {0956540X}
}

@article{Paszke2017AutomaticPytorch,
    title = {{Automatic differentiation in pytorch}},
    year = {2017},
    journal = {31st Conference on Neural Information Processing Systems},
    author = {Paszke, Adam and Gross, Sam and Chintala, Soumith and Chanan, Gregory and Yang, Edward and DeVito, Zachary and Lin, Zeming and Desmaison, Alban and Antiga, Luca and Lerer, Adam}
}

@article{Lecun1998Gradient-basedRecognition,
    title = {{Gradient-based learning applied to document recognition}},
    year = {1998},
    journal = {Proceedings of the IEEE},
    author = {Lecun, Y. and Bottou, L. and Bengio, Y. and Haffner, P.},
    number = {11},
    pages = {2278--2324},
    volume = {86},
    doi = {10.1109/5.726791},
    issn = {00189219}
}

@article{Spall1992MultivariateApproximation,
    title = {{Multivariate stochastic approximation using a simultaneous perturbation gradient approximation}},
    year = {1992},
    journal = {IEEE Transactions on Automatic Control},
    author = {Spall, J.C.},
    number = {3},
    month = {3},
    pages = {332--341},
    volume = {37},
    doi = {10.1109/9.119632},
    issn = {00189286}
}

@article{Griewank1989OnDifferentiation,
    title = {{On automatic differentiation}},
    year = {1989},
    journal = {Mathematical Programming: recent developments and applications},
    author = {Griewank, Andreas and {others}},
    number = {6},
    pages = {83--107},
    volume = {6},
    publisher = {Citeseer}
}

@article{Benedetti2019ParameterizedModels,
    title = {{Parameterized quantum circuits as machine learning models}},
    year = {2019},
    journal = {Quantum Science and Technology},
    author = {Benedetti, Marcello and Lloyd, Erika and Sack, Stefan and Fiorentini, Mattia},
    number = {4},
    month = {11},
    pages = {043001},
    volume = {4},
    doi = {10.1088/2058-9565/ab4eb5},
    issn = {2058-9565}
}

@article{Mitarai2018QuantumLearning,
    title = {{Quantum circuit learning}},
    year = {2018},
    journal = {Physical Review A},
    author = {Mitarai, K and Negoro, M and Kitagawa, M and Fujii, K},
    number = {3},
    month = {11},
    pages = {32309},
    volume = {98},
    doi = {10.1103/PhysRevA.98.032309},
    issn = {2469-9926}
}

@article{Henderson2020QuanvolutionalCircuits,
    title = {{Quanvolutional neural networks: powering image recognition with quantum circuits}},
    year = {2020},
    journal = {Quantum Machine Intelligence},
    author = {Henderson, Maxwell and Shakya, Samriddhi and Pradhan, Shashindra and Cook, Tristan},
    number = {1},
    month = {6},
    pages = {2},
    volume = {2},
    doi = {10.1007/s42484-020-00012-y},
    issn = {2524-4906}
}

@article{Gacon2021SimultaneousInformation,
    title = {{Simultaneous Perturbation Stochastic Approximation of the Quantum Fisher Information}},
    year = {2021},
    journal = {Quantum},
    author = {Gacon, Julien and Zoufal, Christa and Carleo, Giuseppe and Woerner, Stefan},
    month = {10},
    pages = {567},
    volume = {5},
    doi = {10.22331/q-2021-10-20-567},
    issn = {2521-327X}
}

@article{Ostaszewski2019StructureCircuits,
    title = {{Structure optimization for parameterized quantum circuits}},
    year = {2019},
    author = {Ostaszewski, Mateusz and Grant, Edward and Benedetti, Marcello},
    month = {11},
    url = {http://arxiv.org/abs/1905.09692 http://dx.doi.org/10.22331/q-2021-01-28-391},
    doi = {10.22331/q-2021-01-28-391}
}

@article{Shepherd2009TemporallyComputation,
    title = {{Temporally unstructured quantum computation}},
    year = {2009},
    journal = {Proceedings of the Royal Society A: Mathematical, Physical and Engineering Sciences},
    author = {Shepherd, Dan and Bremner, Michael J.},
    number = {2105},
    month = {5},
    pages = {1413--1439},
    volume = {465},
    doi = {10.1098/rspa.2008.0443},
    issn = {1364-5021}
}

@article{McClean2016TheAlgorithms,
    title = {{The theory of variational hybrid quantum-classical algorithms}},
    year = {2016},
    journal = {New Journal of Physics},
    author = {McClean, Jarrod R and Romero, Jonathan and Babbush, Ryan and Aspuru-Guzik, Alán},
    number = {2},
    month = {2},
    pages = {023023},
    volume = {18},
    doi = {10.1088/1367-2630/18/2/023023},
    issn = {1367-2630}
}

@article{Cerezo2021VariationalAlgorithms,
    title = {{Variational quantum algorithms}},
    year = {2021},
    journal = {Nature Reviews Physics},
    author = {Cerezo, M. and Arrasmith, Andrew and Babbush, Ryan and Benjamin, Simon C. and Endo, Suguru and Fujii, Keisuke and McClean, Jarrod R. and Mitarai, Kosuke and Yuan, Xiao and Cincio, Lukasz and Coles, Patrick J.},
    number = {9},
    month = {9},
    pages = {625--644},
    volume = {3},
    doi = {10.1038/s42254-021-00348-9},
    issn = {2522-5820}
}

@article{Luo2020Yao.jl:Design,
    title = {{Yao.jl: Extensible, Efficient Framework for Quantum Algorithm Design}},
    year = {2020},
    journal = {Quantum},
    author = {Luo, Xiu-Zhe and Liu, Jin-Guo and Zhang, Pan and Wang, Lei},
    month = {10},
    pages = {341},
    volume = {4},
    doi = {10.22331/q-2020-10-11-341},
    issn = {2521-327X}
}

\newpage
\section*{Appendix}
\label{sec:appendix}

When comparing SPSB to the parameter shift rule, it makes the most sense to compare in terms of the number of circuit evaluations. However, because the number of parameter updates made by SPSB for a given run is much greater than that of the parameter shift rule, the original data is rather difficult to read. In the main report we present the data with a rolling average. For transparency, the original data is reproduced here.

\begin{figure}
    \captionsetup[subfigure]{justification=justified,margin=0.5\columnwidth,width=5cm,singlelinecheck=false}
    \begin{subfigure}{0.43\textwidth}
    \includegraphics{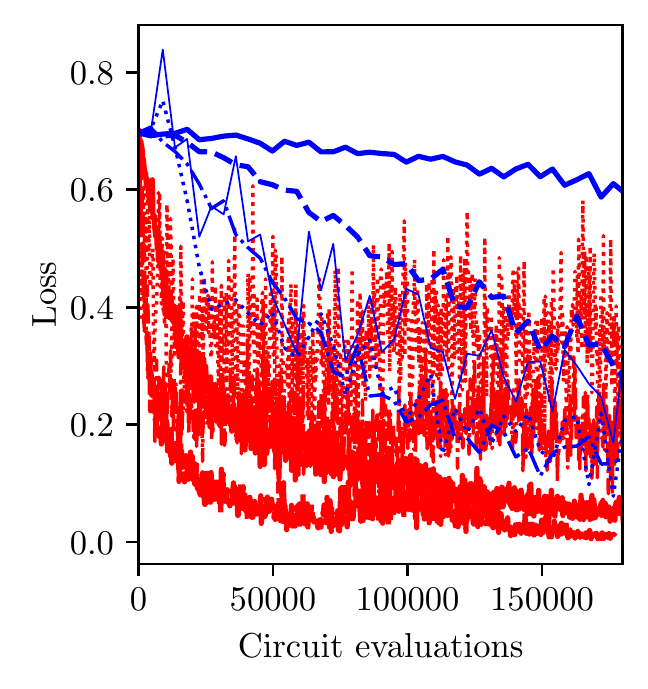}
        \caption{Random data task, $N_B=25$}
        \label{fig:app:lrs:a}
    \end{subfigure}
    \captionsetup[subfigure]{justification=justified,margin=0.8cm,singlelinecheck=false}
    \begin{subfigure}{0.57\textwidth}
        \includegraphics{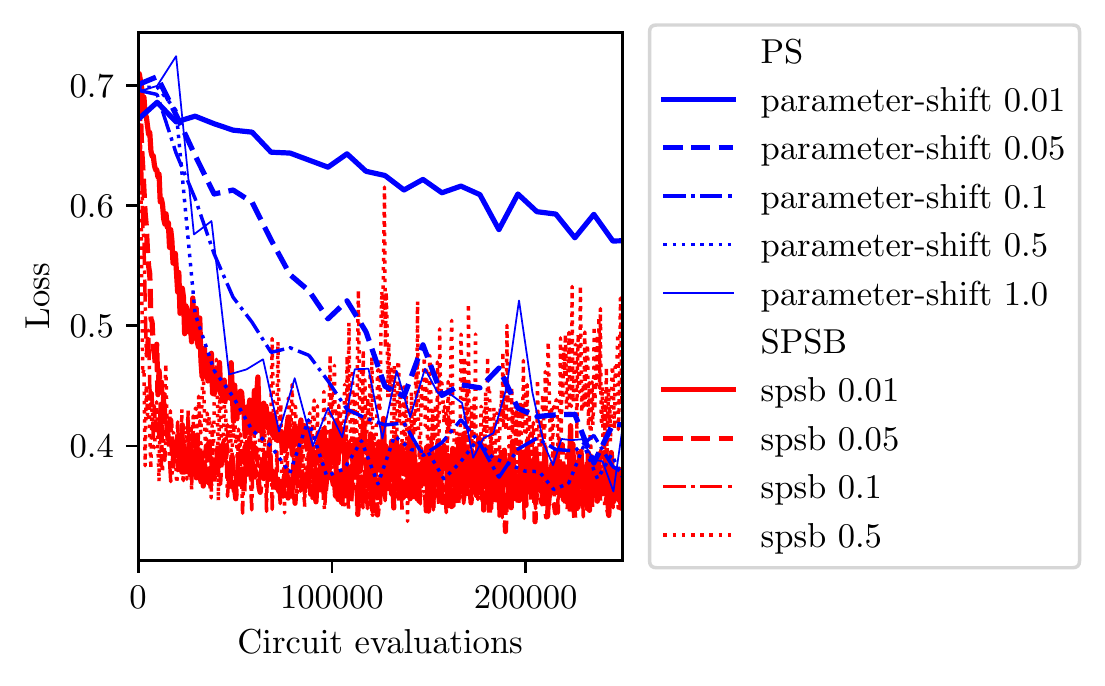}
        \caption{`Quanvolutional' MNIST task, $N_B=50$}
        \label{fig:app:lrs:b}
    \end{subfigure}
    \caption{Figures showing training loss for the parameter shift rule for two different hybrid tasks and a range of learning rates. Figure (a) shows the random datapoints task with 15 features and a batch size of 25. Each line is the median of 10 iterations. Figure (b) shows the quanvolutional network with 4 qubits, as described above, and a batch size of 50. Each line is the median of 5 iterations. To enhance readability of the graph, we only show the loss after every 10th optimisation step for SPSB and every 3rd step for the parameter shift rule.}
    \label{fig:app:learning_rates}
\end{figure}
\end{document}